%% file: MusicPerformanceAnalysis.tex
\documentclass{article}
\usepackage[T1]{fontenc} % add special characters (e.g., umlaute)
\usepackage[utf8]{inputenc} % set utf-8 as default input encoding
\usepackage{ismir,amsmath,cite,url}
\usepackage{graphicx}
\usepackage{color}
\usepackage[flushleft]{paralist}
\setdefaultleftmargin{.9em}{}{}{}{.5em}{.5em}

\usepackage[bookmarks=false,hidelinks]{hyperref}
\usepackage{microtype}
\newcommand{\todo}[1]{{#1}}
\newcommand{\edited}[1]{{#1}}

\newcounter{addcmmts}

\title{Music Performance Analysis: A Survey}

\multauthor
{Alexander Lerch \hspace{1cm}Claire Arthur \hspace{1cm}Ashis Pati \hspace{1cm}Siddharth Gururani} {
Center for Music Technology, Georgia Institute of Technology, Atlanta, USA\\
{\tt\small \{alexander.lerch,claire.arthur,ashis.pati,siddgururani\}@gatech.edu}
}
\def\authorname{Alexander Lerch, Claire Arthur, Ashis Pati, Siddharth Gururani}

\begin{document}

\maketitle
\begin{abstract}
%Music Information Retrieval (MIR) as a research area tends to focus on the analysis of audio signals, but less emphasis is placed on the performance aspect of musical recordings, which is distinct in many ways from the representation of a “song,” “piece,” or musical score. 
Music Information Retrieval (MIR) tends to focus on the analysis of audio signals. Often, a single music recording is used as representative of a ``song'' even though different performances of the same song may reveal different properties. A performance is distinct in many ways from a (arguably more abstract) representation of a ``song,'' ``piece,'' or musical score. The characteristics of the (recorded) performance~---as opposed to the score or musical idea---~can have a major impact on how a listener perceives music. The analysis of music performance, however, has been traditionally only a peripheral topic for the MIR research community. This paper surveys the field of Music Performance Analysis (MPA) from various perspectives, discusses its significance to the field of MIR, and points out opportunities for future research in this field. 
%	\todo{We conclude that while advances in MIR have helped improve MPA research significantly, there is potential for MPA to contribute to MIR research in a more profound manner, and that greater attention to interdisciplinary research would bolster existing approaches in MIR.}
	%\claire{\pcount can we tone down this last sentence? In the end we don't actually discuss much this ``potential''}
\end{abstract}
\section{Introduction}\label{sec:introduction}
    Music, as a performing art, requires a performer or group of performers to render a musical score into an acoustic realization \cite{hill_score_2002}. This is also true for non-classical music: for example, the `score' might be a lead sheet or only a structured sequence of musical ideas, a `performer' could also be a computer rendering audio, and the acoustic realization might be represented by a recording. 
    The performance %~\edited{---at the center of the chain of communication from composer to listener \cite{kendall_communication_1990}---}~
		plays a major role in how listeners perceive a piece of music: even if the score content is identical for different renditions, as is the case in western classical music, listeners may prefer one performance over another and appreciate different `interpretations' of the same piece of music. These differences are the result of the performers actively or subconsciously interpreting, modifying, adding to, and dismissing score information. 
%    It is important to note that this interpretation is desired as the ``constant re-interpretation of music representations is the artistic breath that gives music life'' \cite{lerch_software-based_2009}.\alex{remove this sentence?}

    Although the distinction between score and performance parameters is less obvious for other genres of western music, especially ones without clear separation between the composer and the performer, the concept of interpreting an underlying score is still very much present, be it as a live interpretation of a studio recording or a cover version of another artist's song. In these cases, the freedom of the performers' in modifying the score information is often much higher than it is for classical music~---~reinterpreting a jazz standard can, e.g., include the modification of content related to pitch, harmony, and rhythm.
    
    Performance parameters can have a major impact on a listener's perception of the music \cite{clarke_listening_2002}. Formally, performance parameters can be structured in the same basic categories that we use to describe audio in general: \edited{tempo and timing, dynamics, pitch, timbre} \cite{lerch_introduction_2012}. While the importance of different parameters might vary from genre to genre, the following list introduces some mostly genre-agnostic examples to clarify the performance parameter categories: %\ashis{\pcount Ashis: Can we change the order here to the same as the list in the paragraph or vice versa. Also, }
    \begin{compactitem}
        \item   
        \textit{Tempo and Timing}~---~the score specifies the general rhythmic content, just as it often contains a tempo indicator. While the rhythm is often only slightly modified by performers in terms of micro-timing, the tempo (both in terms of overall tempo as well as expressive tempo variation) is frequently seen only as a suggestion to the performer.
        \item   
        \textit{Dynamics}~---~in most cases, score information on dynamics is missing or only roughly defined. The performers will vary loudness based on their plan for phrasing, tension, importance of certain parts of the score, and highlight specific events with accents.
        \item   
        \textit{Pitch}~---~the score usually defines the general pitches to play, but pitch-based performance parameters include expressive techniques such as vibrato as well as conscious or unconscious choices for intonation.
        \item   
        \textit{Timbre}~---~as the least specific category of musical sound, scores encode timbre parameters often only implicitly (e.g., instrumentation) while performers can, e.g., change playing techniques or the choice of specific instrument configurations (such as the choice of organ registers).
    \end{compactitem}
    Note that usually the performance to be analyzed is a \textit{recording} and not a live performance; every recording contains processing choices and interventions by sound engineer and editor with potential impact on expressivity~---~these modifications cannot be separated from the musicians' creation and are thus an integral part of what is investigated \cite{maempel_musikaufnahmen_2011}. 
    
    The most intuitive form of Music Performance Analysis (MPA)~---discussing, criticizing, and assessing a performance after a concert---~has arguably taken place
    since music was first performed. Early attempts at systematic and empirical MPA can be traced back to the 1930s with vibrato and singing analysis by Seashore
    \cite{seashore_psychology_1938} and the examination of piano rolls by Hartmann \cite{hartmann_untersuchungen_1932}. \edited{In the past two decades, MPA has greatly
    benefited from the advances in audio analysis made by members of the Music Information Retrieval (MIR) community, significantly extending the volume of empirical data by simplifying access to a continuously growing heritage of commercial audio recordings.}
    %\claire{The above sentence has a lot of weird grammar. Here is my suggested rewrite: ``In the past two decades, MPA has greatly benefited from the advances in audio    analysis made by members of the Music Information Retrieval (MIR) community, significantly extending the volume of empirical data thanks in part to a continuously    growing heritage of commercial audio recordings.''} 
    However, while advances in audio content analysis have had clear impact on MPA, the opposite is less true. While there have been publications on performance analysis at ISMIR, the major MIR conference, their absolute number remains comparably small (compare \cite{toyoda_utility_2004,takeda_rhythm_2004,chuan_dynamic_2007,sapp_comparative_2007,sapp_hybrid_2008,liem_expressive_2011,devaney_study_2012,jure_pitch_2012,van_herwaarden_predicting_2014,liem_comparative_2015,bantula_jazz_2016,peperkamp_formalization_2017} with a title referring to music performance out of approximately 1800 ISMIR papers overall).

     %\alex{clarify}
    Historically, MIR researchers often do not distinguish between score-like information and performance information even if the research deals with audio recordings of performances. For instance, the goal of music transcription, a very popular MIR task, is usually to transcribe all pitches with their onset times \cite{Benetos2013automatic}; that means that a successful transcription system transcribes two renditions of the same piece of music differently, although the ultimate goal is to detect the same score (note that this is not necessarily true for all genres).
    %\comment{I would argue that it's often not the goal, especially for non-score based styles where the performance IS the score. That is, where there never is/was any score, the performance is the only available ground truth.}
    %
    Therefore, we can identify a disconnect between MIR research and performance researchers that impedes both the evolution of MPA approaches and robust MIR algorithms, slows gaining new insights into music aesthetics, and hampers the development of practical applications such as new educational tools for music practice and assessment.

\edited{The remainder of this paper is structured as follows. The next} Sect.~\ref{sec:descript} presents research on the objective description and visualization of the performance itself, identifying commonalities and differences between performances. The subsequent sections focus on studies taking these objective performance parameters and relating them to either the performer (Sect.~\ref{sec:performer}) or the listener (Sects.~\ref{sec:listener} and \ref{sec:assessment}). We conclude our overview with a summary on applications of MPA and final remarks in Sect.~\ref{sec:apps}. \edited{Note that while performance research has been inclusive of various musical genres, such as the Jingju music of the Beijing opera \cite{zhang_understanding_2017,rong_gong_automatic_2018}, traditional Indian music \cite{clayton_time_2008,gupta_objective_2012,narang_acoustic_2017} and jazz music \cite{abesser_score-informed_2017}, the vast majority of studies have been concerned with Western classical music. Therefore, the remainder of the paper focuses primarily on Western music.}

%\alex{add applications} 
    %Focus on ISMIR publications (for edits):
    %\begin{itemize}
        %\item   Extraction of Timing deviations from MIDI \cite{toyoda_utility_2004}
								%\item   Tempo curve and tempo model \cite{takeda_rhythm_2004}
        %\item   Extract phrase boundaries with performance info \cite{chuan_dynamic_2007}
        %\item   piano: tempo, dynamics, visualization \cite{sapp_comparative_2007,sapp_hybrid_2008}
        %\item   tempo/timing individuality \cite{liem_expressive_2011}
        %%\item   intonation of three-part singing \cite{devaney_study_2012}
        %\item   pitch visualization \cite{jure_pitch_2012}
        %\item   trained NN for piano performance \cite{van_herwaarden_predicting_2014}
        %\item   orchestra performance clustering \cite{liem_comparative_2015}
        %\item   data-driven jazz performance modeling \cite{bantula_jazz_2016}
        %\item   represent local tempo variations \cite{peperkamp_formalization_2017}
    %\end{itemize}

\section{\todo{Performance measurement}}\label{sec:descript}
%\ashis{\pcount Ashis: rename section heading to Descriptive MPA?}
%\claire{Gabrielsson calls it ``performance measurement'' in his 2003 review. Perhaps adopt?}
% Introductory para
A large body of work focuses on a descriptive approach to analyzing performance recordings. 
Such studies typically extract characteristics such as the tempo curve \cite{repp_patterns_1990, palmer_mapping_1989, povel1977temporal} or loudness curve \cite{repp_microcosm_1998, seashore_psychology_1938} from the audio and aim at either gaining general knowledge on performances or comparing attributes between different performances/performers based on trends observed in the extracted data.

% Tempo related stuff
Several researchers observed a close relationship between musical phrase structure and deviations in tempo and timing \cite{shaffer_timing_1984, povel1977temporal, palmer_music_1997}. 
For example, tempo changes in the form of \textit{ritardandi} tend to occur at phrase boundaries \cite{lerch_software-based_2009, palmer_mapping_1989}. 
A similar co-occurrence was observed between dynamic patterns and timing \cite{repp_dynamics_1996,lerch_software-based_2009}. 
Additionally, Dalla Bella found the overall tempo influences the overall loudness of a performance \cite{dalla_bella_tempo_2004}. 
There are also indications that loudness can be linked to pitch height \cite{repp_dynamics_1996}. 
While the close relation of tempo and dynamics to structure has been repeatedly verified, Lerch did not succeed in finding similar relationships between structure and timbre properties in the case of string quartet recordings \cite{lerch_software-based_2009}.

Pitch-based performance parameters have been analyzed mostly in the context of single-voiced instruments. Vibrato and its rate has, e.g., been studied for vocalists \cite{seashore_psychology_1938,devaney_automatically_2011} and violinists \cite{macleod2006influences, dimov_short_2010}. Regarding intonation, Devaney et al.\ found significant differences between professional and non-professional vocalists in terms of the size of the interval between semi-tones \cite{devaney_automatically_2011}.

% Instrument-specific stuff
%Many researchers tend to focus on specific instruments. The piano, in particular, has received a lot of attention. 
%Repp studied the relationship between pedaling and the overall tempo of a piano performance \cite{repp_pedal_1996, repp_effect_1997}. 
%Goebl et al.\ studied synchronicity between the left and right hand or the melody and accompaniment in piano performance, respectively, identifying a variety of between-hand asynchronies that were also easily predicted with simple probabilistic models \cite{goebl_investigations_2010}. 
%With respect to non-piano performances, Devaney et al.\ studied automatic methods to extract note timings, intonation, vibrato rates, and dynamics from vocal performances \cite{devaney_automatically_2011}. 
%They found significant differences between professional and non-professional vocalists in terms of the size of the interval between semi-tones.
%Several other authors have studied vibrato rates and width in violin performances \cite{macleod2006influences, dimov_short_2010}. 

Other studies use a multitude of performance parameters and aim to identify trends over time. For example, Ornoy and Cohen investigated violin performances of $19$th century repertoire recorded in the past two decades \cite{ornoy2018analysis}. 
They found a blend of stylistic approaches among  violinists which questions the traditional distinction made between a historically informed and a mainstream performance. 

The challenges in accessibility and interpretability of the extracted performance parameters have also led researchers to work on more intuitive or condensed forms of visualization that allow describing and comparing different performances beyond the traditional forms of visualization such as tempo curves \cite{repp_patterns_1990, palmer_mapping_1989, povel1977temporal} and scatter plots \cite{lerch_software-based_2009}.
%
%Visualization of performance parameters is another aspect that allows describing and comparing performances, possibly allowing a more intuitive understanding of the data. 
%While visualizations such as tempo curves \cite{repp_patterns_1990, palmer_mapping_1989, povel1977temporal} and scatter plots \cite{lerch_software-based_2009} are common, some specialized visualizations have been proposed. 
The ``performance worm,'' for example, is a pseudo-3D visualization of the tempo-loudness space that allows the identification of specific gestures in that space  \cite{langner_representing_2002,dixon_performance_2002}. 
Sapp proposed so-called ``Timescapes'' and ``Dynascapes'' to visualize subsequence similarities of performance parameters \cite{sapp_comparative_2007,sapp_hybrid_2008}.
%\claire{\pcount ``subsequent'' is a bit ambiguous here..}

% Computational studies
While most of the studies mentioned above make use of statistical methods to extract, visualize, and investigate patterns in the performance data, few studies make use of Machine Learning (ML) for MPA and performance modeling. 
However, \edited{ML-based approaches are useful for tasks such as composer classification,} discovery of performance rules, or modeling performance characteristics. 
%\claire{the above sentence was ambiguous. I adjusted it to what I best thought you were trying to say.}
Widmer conducted studies that utilized ML to model expression in musical performance \cite{widmer_computational_2004,widmer_automatic_2004, Widmer97applicationsof} \edited{and to learn simple rules from performance data with inductive learning \cite{widmer_discovering_2003}. 
%\claire{\pcount music theory is an entire discipline.. can we be more specific here? Like, ``incorporating phrase-boundaries as defined by rules of music theory'' or whatever}
He} also applied ML to identify performers, showing that performer characteristics can be modeled by ML algorithms \cite{widmer_automatic_2004}. 
%Additionally, ML has also been used for describing musical performances in terms of the playing techniques used, e.g., identifying playing techniques in drum recordings \cite{wu_drum_2016}.
%\claire{\pcount suggestion for last sentence: ``Additionally, ML has been used for describing musical performances in terms of the playing techniques used, such as (fill appropriate drumming techniques here) in drum recordings''. Just a bit weird to use ``playing techniques'' twice in close succession.}

The studies presented in this section often follow an exploratory approach; extracting various parameters in order to identify commonalities or differences between performances. 
While this is, of course, of considerable interest, one of the main challenges is the interpretability of these results. 
Just because there is a timing difference between two performances does not necessarily mean that this difference is perceptually meaningful. 
Without this link, however, results can only provide limited insights into which parameters and parameter variations are ultimately important. 
Another typical challenge in such studies is the reliability of MIR algorithms for automatic annotations. 
\edited{While the accuracy of such algorithms has steadily improved over time, the fact that the majority of studies surveyed here continue to rely on manually-annotated data implies that the state-of-the-art algorithms for automatic annotation still lack the required degree of accuracy for most tasks.}%While the accuracy of such algorithms has steadily improved over time, the use of manually annotated data in the majority of presented studies indicates that the current state-of-the-art in MIR lacks the required degree of accuracy or robustness. %\comment{does this sound okay? The next sentence is what it used to be.}
%While the accuracy of such algorithms has steadily improved over time, the use of manually annotated data in the majority of the presented studies indicates high requirements on accuracy and robustness that current MIR algorithms fail to fulfill.
%\claire{\pcount this is still weird, sorry! Suggested rewrite: ``While the accuracy of such algorithms has steadily improved over time, the fact that the majority of studies surveyed here continue to rely on manually-annotated data implies that the state-of-the-art algorithms for automatic annotation still lack the required degree of accuracy for most tasks.}
%This leads to researchers having to manually annotate or correct computer-generated annotations. 
As a result, most analyses are performed on small sample sizes possibly leading to poor generalizability of the studies.
%\ashis{\pcount Ashis: possibly end this with a line on how the MIR community can help in this regard?}

% A typical challenge is also that the used sample sizes are small, questioning the generalizability and significance of results.

% \alex{use of MIR vs. manually annotated data, scalability}    

\section{\todo{Performer}}\label{sec:performer}
%\ashis{\pcount Ashis: Rename section title to Performer-focussed MPA?}
    While most studies focus on the extraction of performance parameters or the mapping of these parameters to the listeners' reception (see Sects.~\ref{sec:listener} and \ref{sec:assessment}), some investigate the capabilities, goals, and strategies of performers. 
    A performance is usually based on an explicit or implicit performance plan with clear intentions \cite{clarke_understanding_2002}. There is, as Palmer verified, a clear relation between reported intentions and objective parameters related to phrasing and timing of the performance \cite{palmer_mapping_1989}. 
    %\claire{\pcount it's a bit weird that the above sentence implies that you're going to say what Palmer found but then cites Juslin.} 
    \edited{Similar} relations between the intended emotionality and loudness and timing measures were reported by Juslin \cite{juslin_cue_2000} and Dillon
    \cite{dillon_extracting_2001,dillon_statistical_2003,dillon_recognition_2004}. For example, \todo{projected} %\claire{do you mean ``implied'' here? We were talking about performer intentions and measured parameters} 
		emotions such as anger and sadness show significant correlations with high and low tempo and high and low overall sound level, respectively.
    %\claire{\pcount is this really at performer level? This seems like something that would be indicated in the score. For instance, Huron found exactly this kind of thing doing computational work with symbolic data alone.}
    Moreover, a performer's control of expressive variation has been shown to significantly improve the conveyance of emotion. For  instance,  a  study  by  Vieillard  et  al.\ found that listeners were better able to perceive the presence  of  specific  emotions in  music  when  the  performer played an expressive (versus mechanical) rendition of the composition \cite{vieillard_expressiveness_2012}. 
    This suggests that the performer plays a fairly large role in communicating an emotional ``message'' above and beyond what is communicated through the score alone \cite{juslin_communication_2003}.
    In addition to the performance plan itself, there are other \todo{influences} %\comment{Sidd: do you mean ``factors''??} 
		shaping the performance. Acoustic parameters of concert halls such as the early decay time have been shown to impact performance parameters such as tempo \cite{scharer_kalkandjiev_influence_2013,scharer_kalkandjiev_influence_2015,luizard_singing_2019}. 
		%Possibly somewhat related are the findings by Repp that pedaling in piano performance is related to the overall tempo  \cite{repp_pedal_1996, repp_effect_1997}.
    %\claire{\pcount last sentence necessary?}
    Another interesting area of research is performer error. Repp analyzed performers' mistakes and found that errors were concentrated in mostly unimportant parts of the
    score (e.g., middle voices) where they are harder to recognize \cite{huron_tone_2001}, suggesting that performers consciously or unconsciously avoid salient mistakes \cite{repp_art_1996}. 

            %\alex{\pcount should we remove the paragraph for space reasons?}\ashis{Ashis: possibly}
	    %\claire{I'm fine with that}
            %Other studies investigate the importance of the feedback of the music instrument to the performer \cite{sloboda_music_1982}; there have been studies reporting on the effect of deprivation of auditory feedback \cite{repp_effects_1999}, investigating the performers' reaction to delayed or changed auditory feedback \cite{pfordresher_effects_2002, finney_auditory_2003, pfordresher_auditory_2005}, or evaluating the role of tactile feedback in a piano \hbox{performance \cite{goebl_tactile_2008}}.

    There is a wealth of information about performances that can be learned from performers. 
    The main challenge of this direction of inquiry is that such studies \edited{have to involve} the performers themselves. 
    %\claire{\pcount I don't really get this. Do you mean like interviewing them? I think being more specific would help clarify the problem}
    This limits the amount of available data and possibly excludes well-known and famous artists, resulting in a possible lack of generalizability.
    Depending on the experimental design, the separation of possible confounding variables (e.g., motor skills, random variations, and the influence of common performance rules) from the scrutinized performance data can be a challenge.

% LISTENER SECTION
\input{sections/listener.tex}

\section{\todo{Performance assessment}}\label{sec:assessment}
%\ashis{\pcount Ashis: rename title to Performance-focussed MPA}
    Assessment-focused MPA deals with modeling how we as humans assess a musical performance. %While this is technically a specific case of listener-focused MPA, it is    of central importance due to its role in music education. 
		\edited{While this is technically a     subset of listener-focused MPA, its importance to MPA research and music education warrants a tailored review of research in this area.} Performance assessment is a critical and ubiquitous aspect of music pedagogy: students rely on regular feedback from teachers to learn and improve skills, recitals are used to monitor progress, and selection into ensembles is managed through competitive auditions.
    The performance parameters on which these assessments are based are not only subjective but also ill-defined, leading to large differences in subjective opinion among music educators \cite{thompson2003evaluating,wesolowski2016examining}. %Thus, there is a need to formalize the rubrics and procedures for such assessments. 
    Work within Assessment-focused MPA seeks to increase the objectivity of performance assessments \cite{mcpherson_assessing_1998}, and build accessible and reliable tools for automatic assessment.

    Over the last decade, several researchers have worked towards developing tools capable of automatic music performance assessment which can be categorized based on:
    \begin{inparaenum}[(i)] 
        \item the parameters of the performance that are assessed, and
        \item the technique/method used to design these systems.
    \end{inparaenum}
    
    %\subsection{Assessment Parameters}
    Tools for performance assessment typically assess one or more performance parameters which are usually related to the accuracy of the performance in terms of pitch and timing \cite{wu_towards_2016,vidwans_objective_2017,pati_assessment_2018,luo_detection_2015}, or quality of sound (timbre) \cite{knight_potential_2011,picas_real-time_2015}. %Claire: an example of what "quality of sound" means might be helpful.
    %Some tools may also focus on the number of mistakes %in performances
    %Claire: how is this different from detecting accuracy? Is this necessary?
    %\cite{luo_detection_2015}. 
    In building an assessment tool, the choice of parameters may depend on the proficiency level of the performer being assessed. For example, beginners will benefit more from feedback in terms of low-level parameters such as pitch or rhythmic accuracy as opposed to feedback on higher-level parameters such as articulation or expression.

    Assessment tools can also vary based on the granularity of assessments. Tools may simply classify a performance as `good' or `bad' \cite{knight_potential_2011, nakano2006automatic}, or grade it on a scale, say from $1$ to $10$ \cite{pati_assessment_2018}. %Temporal granularity is also an important design choice.
    %Claire: remove prior sentence?
    Systems may provide fine-grained note-by-note assessments \cite{picas_real-time_2015} or analyze entire performances and report a single assessment score \cite{nakano2006automatic,pati_assessment_2018}. 
    % The performance parameter being assessed will play a critical role in this decision: while fine-grained assessments for low-level performance parameters such as note accuracy are possible, it may not be possible for higher-level parameters such as expressivity. 
    
    %\subsection{Assessment Methods}
    While different methods have been used to create performance assessment tools, the common approach has been to use descriptive features extracted from the audio recording of a performance based on which a cognitive model predicts the assessment. This approach requires availability of performance data (recordings) along-with human (expert) assessments for the rated parameters. 

    The level of sophistication of cognitive models was limited especially for early attempts; e.g., simple classifiers such as Support Vector Machines were used to predict human ratings. In this case, descriptive features became an important aspect of the system design. 
    %Claire: descriptive features were not important for cognitive models? I don't understand that last sentence.
    In some approaches, standard spectral and temporal features such as Spectral Centroid, Spectral Flux, and Zero-Crossing Rate were used \cite{knight_potential_2011}. In others, \edited{features aimed at capturing cognitive aspects of music perception} were hand-designed using either musical intuition or expert knowledge 
    %\claire{\pcount how do we know they are cognitively important? Can we simply ditch the ``cognitively important'' part? You may have to adjust the ``cognitively intuitive'' line below.}
    \cite{nakano2006automatic,abesser2013automatic,picas_real-time_2015,li_analysis_2015}. For instance, Nakano et al.\ used features measuring pitch stability and vibrato as inputs to a simple classifier to rate the quality of vocal performances \cite{nakano2006automatic}. Several studies also attempted to combine low-level audio features with hand-designed feature sets \cite{luo_detection_2015,wu_towards_2016,vidwans_objective_2017}, as well as incorporating information from the musical score into feature computation \cite{devaney_study_2012,bonada2009performance,vidwans_objective_2017,bozkurt2017voice}. 
    
    Recent methods, however, have transitioned towards using advanced ML techniques such as Sparse Coding \cite{han_hierarchical_2014,wu_learned_2018,wu_assessment_2018} and Deep Learning \cite{pati_assessment_2018} as a proxy to sophisticated cognitive models. 
    % Some recent works use techniques such as Sparse Coding \cite{han_hierarchical_2014,wu_learned_2018,wu_assessment_2018} and Deep Learning \cite{pati_assessment_2018}. 
    Contrary to earlier methods which focused on extracting cognitively intuitive or important features, these techniques input raw data (usually in the form of pitch contours or spectrograms) and train the models to automatically learn meaningful features so as to accurately predict the assessment ratings.  
    
    In some ways, this evolution in methodology has mirrored that of other MIR tasks: there has been a gradual transition from feature design to feature learning. Feature design and feature learning have an inherent trade-off. Learned features extract relevant information from data which might not be represented in the hand-crafted feature set. This is evident from their superior performance at assessment modeling tasks \cite{wu_assessment_2018,pati_assessment_2018}. However, this superior performance comes at the cost of low interpretability. Learned features tend to be abstract and cannot be easily understood. Custom-designed features, on the other hand, typically either measure a simple low-level characteristic of the audio signal or link to high-level semantic concepts such as pitch or rhythm which are intuitively interpretable. Thus, such models allow analysis that can aid in the interpretation of semantic concepts for music performance assessment. For instance, Gururani et al.~analyzed the impact of different features on an assessment prediction task and found that
    \begin{inparaitem}[]
        \item features measuring tempo variations were particularly critical, and that 
        \item score-aligned features performed better than score-independent features \cite{gururani_analysis_2018}.
    \end{inparaitem}
    
    %\subsection{Challenges}
    In spite of several attempts across varied performance parameters using different methods, the important features for assessing music performances remain unclear. This is evident from the average performance of these tools in modeling human judgments. Most of the presented systems either work well only for very select data \cite{knight_potential_2011} or have comparably low prediction accuracies \cite{vidwans_objective_2017, wu_towards_2016}, rendering them unusable in most practical scenarios. While this may be partially attributed to the subjective nature of the task itself, there are several other factors which have limited the improvement of these tools. 
    First, most of the models are trained on small task-specific or instrument-specific datasets that might not reflect noisy real-world data. 
    This reduces the generalizability of these models. The problem becomes more serious for data-hungry methods such as Deep Learning which require large amounts of data for training. 
    Second, the distribution of ground-truth (expert) ratings given by human judges is in many datasets skewed towards a particular class or value \cite{gururani_analysis_2018}. This makes it challenging to train unbiased models.
    Finally, the number of parameters required to adequately model a performance results in high dimensional data. While the typical approach is to train different models for different parameters, this approach necessitates availability of performance data along-with expert assessments for all these parameters. In many occasions, such assessments are either not available or are costly to obtain. %In addition, the assessment parameters are often highly correlated \cite{pati_assessment_2018}.%Claire: suggest adding why this matters (e.g., causes problems for the model or requires additional preprocessing, etc.)
        
    %In addition to the above limitations which need to be addressed, there are several other challenging problems for MIR researchers interested in this domain. Better techniques need to be developed to factor the score information into the assessments. So far, the primary approach for this has been using dynamic time warping (DTW) based methods \cite{vidwans_objective_2017,bozkurt2017voice} to compute distance-based features between the score and the performance. However, expressive performances are supposed to deviate from the score \comment{[]} and simple distance-based features may fail to adequately capture the nuances. The problem of how to incorporate this information into the assessment computation process remains an open challenge which deserves more attention.\comment{should we remove this paragraph?} 

    Given these data-related challenges, an interesting direction for future research might consider leveraging models which are successful at assessing a few parameters (and/or instruments) to improve the performance of models for other parameters (and/or instruments). This approach, usually referred to as transfer learning, has been found to be successful in other MIR tasks \cite{choi2017transfer}.
    In addition, the ability to interpret and understand the features learned by end-to-end models will play an important role in improving assessment tools. 
		\edited{Interpretability of neural networks is still an active area of research in MIR, and performance assessment is an excellent test-bed for developing such methods.}
		%This is an active area of research in MIR and performance assessment, and is an excellent test-bed on which such techniques can be developed and evaluated. 
    %\claire{\pcount I tried to fix the grammar here by adding a missing ``and'' but it still isn't right. ``such techniques'' needs a modifier and a reference    point.}

    %\section{Current challenges}
%    \subsection{Data}
%        \begin{itemize}
%            \item   data acquisition 
%            \item   data set sizes
%            \item   data set (scoring) distribution
%            \item   noisy labels
%        \end{itemize}

\section{Conclusion}\label{sec:apps}
%The insights gained from MPA are impactful to a large range of applications. 
    %The previous sections outlined insights gained by MPA at the intersection of audio content analysis, empirical musicology, and music perception research. This data
    %\claire{what data?} is of importance for better understanding the process of music making as well as affective user reactions to music. Furthermore, it
    %\claire{what is "it"?} enables a considerable range of applications spanning a multitude of different areas including systematic musicology, music education, MIR, and computational creativity, leading to a new generation of music discovery and recommendation systems, and generative music systems.
    %\claire{\pcount I really hate this opening paragraph to the conclusion! Also, there are too many ambiguous "it" and "this" as subjects to understand what this paragraph is saying.}

\edited{The previous sections outlined insights gained by MPA at the intersection of audio content analysis, empirical musicology, and music perception research.  These insights are of importance for better understanding the process of  making music as well as affective user reactions to music. Furthermore, they enable a considerable range of applications spanning a multitude of different areas including systematic musicology, music education, MIR, and computational creativity, leading to a new generation of music discovery and recommendation systems, and generative music systems.}
    % moved here from introduction   
    %A clear distinction between score-inherent and performance-inherent parameters can lead to better understanding the process of music making as well as affective user reactions to music. Furthermore, it can enable a new generation of targeted applications in music tutoring systems, music discovery, recommendation systems, and generative music systems. Thus, applications span a multitude of different areas including systematic musicology, music education, Music Information Retrieval (MIR), and computational creativity.
%
%
    %
    The most obvious application example connecting MPA and MIR is music tutoring software. Such software aims at supplementing %or replace \ashis{\pcount Ashis: do we want to say replace ?} 
		teachers by providing students with insights and interactive feedback by analyzing and assessing the audio of practice sessions. The ultimate goals of an interactive music tutor are to highlight problematic parts of the students' performance, provide a concise yet easily understandable analysis, give specific and understandable feedback on how to improve, and individualize the curriculum depending on the students' mistakes and general progress. Various (commercial) solutions are already available, exhibiting a similar set of goals. These systems adopt different approaches, ranging from traditional music classroom settings to games targeting a playful learning experience. Examples for tutoring applications are SmartMusic \cite{makemusic_inc._smartmusic_2018}, Yousician \cite{yousician_2019}, Music Prodigy \cite{the_way_of_h_inc._dba_music_prodigy_music_2018}, and SingStar \cite{sony_interactive_entertainment_europe_singstar_2018}.
    %Most of these systems focus on beginner-level students, since software for higher-level students requires the implementation of more advanced knowledge of music performance conventions and rules.

    Performance parameters have a long history being either explicitly or implicitly part of MIR systems. For instance, core MIR tasks such as music genre classification and music recommendation systems have been utilizing tempo and dynamics features successfully \cite{fu_survey_2011}. %Tempo, for instance, is an ubiquitous feature for many such systems, and features representing musical dynamics are standard as well.
    Generative models often require performance data to allow for the rendition of a convincing output. This obviously includes performance rendition systems that take a score and attempt to render a human-like output \cite{malik2017neural,oore2018time}, but it is also important for models of improvisation such as jazz solos as pitch information is part of the performance.
  
    Despite such practical applications, there are still many open topics and challenges that need to be addressed. The main challenges of MPA have been summarized at the end of the sections above. The related challenges to the MIR community, however, are multi-faceted as well.
    First, the fact that the majority of the presented studies use manual annotations instead of automated methods should encourage the MIR community to re-evaluate the measures of success of their proposed systems if, as it appears to be, the outputs lack the robustness or accuracy required for a detailed analysis even for \todo{tasks considered to be 'solved.'} %\ashis{\pcount Ashis: I feel like this needs an example.} 
    Second, the missing separation of score and performance parameters when framing research questions or problem definitions can impact not only interpretability and reusability of insights but might also reduce algorithm performance. If, e.g., a music emotion recognition system does not differentiate between the impact of core musical ideas and performance characteristics, it will have a harder time differentiating relevant and irrelevant information. Thus, it is essential for MIR systems to not only differentiate between score and performance parameters in the system design phase but also analyze their respective contributions during evaluation.
    %Third, \alex{\pcount Hmm. Anybody got a good third point here? Would be nice to have 3... Maybe something about how we can use MPA insights in MIR systems?}
    %\claire{do we have any suggestion on how to improve the small data problem? Meta analyses? Poor generalizability may also have to do with the often misguided
    %assumption that because a model predicted some parameter as being important, that it is perceptually meaningful.}
    \edited{Third, lack of data continues to be a challenge for both, MIR core tasks and MPA; a focus on approaches for limited data \cite{mcfee_software_2015}, weakly labeled data, and unlabeled data \cite{wu_labeled_2018} could help address this problem.}.
		
    %While there is always a call for general improvements of analysis algorithms (e.g., robust transcription of playing techniques, high accuracy of timing extraction, etc.), one issue that can be often observed is the missing separation of score and performance parameters when framing a problem definition; if, e.g., an emotion recognition system does not differentiate between the impact of core musical ideas and performance characteristics, then not only are interpretability and insights impacted but the system might have a harder time differentiating relevant from irrelevant information. Thus, it is essential for MIR systems to not only differentiate between score and performance parameters in the system design phase but also analyze their respective contributions during evaluation.
    
    In conclusion, the fields of MIR and MPA each depend on the advances in the other field. 
    %suggested addition:
    In addition, music perception and cognition, while not a traditional topic within MIR, can be seen as an important linchpin for the advancement of MIR systems that depend on reliable and diverse perceptual data. Cross-disciplinary approaches to MPA bridging methodologies and data from music cognition and MIR are likely to be most influential for future research.
    Empirical, descriptive research driven by advanced audio analysis is necessary to extend our understanding of music and its perception, which in turn will allow us to create better systems for music analysis, music understanding, and music creation.
\bibliography{musicperformance,ismir-performance}

\end{document}

%% file: sections/listener.tex
\section{\todo{Listener}}\label{sec:listener}
%\ashis{\pcount Ashis: rename section title to Listener-focussed MPA?}
    Every performance will ultimately be received and processed by a listener. 
The listener's meaningful interpretation of the incoming musical information relies on a
sophisticated network of parameters. 
These include not only external, or semi-objective parameters such as score or performance-based features, but also ``internal'' ones such as those shaped by the culture, training, and history of the listener. 
For this reason, listener-focused MPA remains one of the most challenging and elusive areas of research. 
However, to the extent that MPA research depends on purely perceptual information (e.g., expressiveness) or intends to deliver perceptually-relevant output (e.g., performance evaluation or reception, similarity), it is imperative to achieve a fuller understanding of the perceptual relevance of the manipulation and interaction of performance characteristics (e.g., tempo, dynamics, articulation). 
%The subsequent paragraphs provide a brief overview of the relevant literature on music perception and MPA, along with some discussion of the relevance of this information for current and future work in both MPA and in MIR in general.\alex{\pcount do we need this last sentence?}\ashis{Ashis: I think it will be abrupt wihtout this sentence.}
    
\subsection{Musical expression}
    %what do listeners care about?
%    \alex{question here: starting with a question is stylistically quite different from the previous sections --- should we adapt or leave it as it is? I don't mind either way, just wanted to bring it up.}
%\claire{changed it.}
When it comes to listener judgments of a performance, it remains poorly understood which aspects are most important, salient, or pertinent for the listener's sense of satisfaction.
According to Schubert and Fabian \cite{schubert_taxonomy_2014}, listeners are very concerned with the   notion of ``expressiveness'' which is a complex, multifaceted construct.
%they care a lot about expression (and emotion). What does "expression" mean, then?
Discovering which performance characteristics contribute to an expressive performance thus requires     dissecting what listeners deem ``expressive'' as well as understanding the relation and potential differences between measured and perceived performance features.
For instance, expressiveness is style-dependent, meaning that the perceived \textit{appropriate} level  of expression in a Baroque piece will be different from that of a Romantic piece~---~something that has     been referred to as ``stylishness'' \cite{fabian_baroque_2009, kendall_communication_1990}.
In addition, there is the perceived \textit{amount} of expressiveness, which is considered independent  of stylishness \cite{schubert_dimensions_2006}.
Finally, Schubert and Fabian distinguish a third ``layer'' of expressiveness which arises from a        performer's manipulation of   various features specifically to alter or enhance emotion \cite{schubert_taxonomy_2014}.
This is distinct from \textit{musical} expressiveness which more generally refers to the manipulation   of compositional elements by the performer in order to be ``expressive'' without necessarily needing to express a specific emotion.
Practically speaking, however, it may be difficult for listeners to separate these varieties of expressiveness  \cite[p.293]{schubert_taxonomy_2014}), and research has demonstrated that there are interactions between them (e.g., \cite{vieillard_expressiveness_2012}).

\subsection{Expressive variation} \label{ssec:expr_var}
    %What parameters are important for communicating "expressivity" or "emotion"? What function does this performance expression serve?
 Several scholars have made significant advances in our understanding of the role of timing, tempo, and dynamics on listeners’ perception of music. 
As noted in Sect.~\ref{sec:descript}, the subtle variations in tempo and dynamics executed by a performer have been shown to play a large role in highlighting and segmenting musical structure. 
For instance, changes in timing and articulation within small structural units such as the measure, beat, or sub-beat appear to aid in the communication and emphasis of the metrical structure (e.g., \cite{sloboda_communication_1983, gabrielsson_once_1987, palmer_timing_1989, behne_musikpsychologische_1993}), whereas changes across larger segments such as phrases, aid in the communication of formal structure.
In fact, the communication of musical structure has been suggested as one of the primary roles or functions of a successful performer (see \cite{repp_obligatory_1998, juslin_five_2003}). 
For instance, an experiment by Sloboda found that listeners were      better able to identify the meter of an ambiguous passage when performed by a more experienced performer \cite{sloboda_communication_1983}.   
Through measuring the differences in the performers' expressive variations, Sloboda identified dynamics and articulation~---in particular, a \emph{tenuto} articulation---~as the most important  features for communicating which notes were accented.
 
The extent to which a performer's expressive variations align with a listener's musical expectations appears an important consideration.
For example, because of the predictable relation between timing   and structural segmentation, it has been demonstrated that listeners find it difficult to detect
timing (and duration) deviations from a ``metronomic'' performance when the pattern and placement of those deviations are stylistically typical \cite{repp_patterns_1990, repp_constraint_1992, ohriner_grouping_2012}.
Likewise, Clarke \cite{clarke_imitating_1993} found pianists able to more accurately reproduce a    performance when the timing profile was ``normative'' with regards to the musical structure, and also found listeners' aesthetic judgments to be highest for those performances with the original timing profiles       compared with those that were inverted or altered.
    
%\ashis{\pcount Ashis: the second para also started with the idea that performance parameters do more than communicate structural information. Hence, the introductory setence of this para feels a little weird to me.} 
In addition to communicating structural information to the listener, the role of performance features such as timing and dynamics have also been studied extensively for their role in shaping “expressive” performance (see \cite{clarke_rhythm_1998, gabrielsson_performance_1998}). For instance, a factor analysis in \cite{schubert_taxonomy_2014} examined the features and qualities that may be related to perceived expressiveness, finding that
dynamics had the highest impact on the factor labeled
``emotional expressiveness.'' Gingras et al.\ studied the relation between musical structure, expressive variation, and listeners' ratings of musical tension. 
They found that variations in expressive timing were most predictive of listeners' tension ratings \cite{gingras_linking_2016}.
While the role of expressive variation in timbre and intonation have generally been less studied, there has been substantial attention given to the expressive qualities of the singing voice  where these parameters are especially relevant (see \cite{fruhholz_singing_2018}). 
For instance, Sundberg        found that a sharpened intonation at a phrase climax contributed to increased      expressiveness and perceived excitement  \cite{sundberg_intonation_2013}, and Siegwart and Scherer found that listener preferences   were correlated with certain spectral components such as the relative strength of the fundamental and       higher spectral centroid \cite{siegwart_acoustic_1995}.
 %more features and their effects here? 
    %For a recent review of listeners' judgments of expressiveness in music performance, see                 \cite{schubert_taxonomy_2014}.
    %\alex{a bit of a sudden end... any conclusions or summarizing comments for    this section?}

The reason why expressive variation is so enjoyable for listeners remains largely an open research question.
As mentioned above, its role appears to go beyond bolstering the communication of musical structure. 
As pointed out by Repp, even a computerized or metronomic performance     will contain grouping cues \cite{repp_obligatory_1998}. 
However, one prominent theory suggests that systematic performance deviations (such as tempo) may generate aesthetically pleasing expressive performances in part due to     their exhibiting characteristics that mimic ``natural motion'' in the physical world                        \cite{gjerdingen_shape_1988, todd_dynamics_1992, repp_music_1993, todd_kinematics_1995, van_noorden_resonance_1999} or human movements or gestures \cite{ohriner_grouping_2012, broze_iii_animacy_2013}.
For instance, Friberg and Sundberg suggested that the shape of final ritardandi matched the the         velocity of runners coming to a stop \cite{friberg_does_1999}, and Juslin includes ``motion principles'' %as the ``M'' in his GERMS 
    in his model of performance expression \cite{juslin_five_2003}.
%\claire{still a sudden ending I guess, but I don't think we have room for filler}

   \subsection{Mapping and Predicting Listener Judgments}
    %previously ``performance characteristics''
    %Methodology: how do we know we are measuring the performance/performer as opposed to the song, piece,  or composition??
%In terms of characteristics, is it interesting to note Repp's 1997 study (recently replicated) that    finds listeners prefer average performer variations in expressive timing and dynamics?
%Put Repp 1997 and replication (2018) here?
In order to isolate listeners' perception of features that are strictly performance-related, several    scholars have investigated listeners' judgments across multiple performances of the same excerpt of music (e.g.,      \cite{repp_patterns_1990,  fabian_musical_2008}). 
A less-common technique relies on synthesized constructions or manipulations of performances, typically   using some kind of rule-based system to manipulate certain musical parameters (e.g.,
    \cite{repp_expressive_1989, sundberg_how_1993, clarke_imitating_1993, repp_obligatory_1998}), and       frequently making use of continuous data collection measures (e.g., \cite{schubert_taxonomy_2014}).
    
From these studies, it appears that listeners (especially ``trained'' listeners) are capable not only   of identifying performance characteristics such as phrasing,
    articulation, and vibrato, but that they are frequently able to identify them in a manner that is aligned with the performer's intentions (e.g., \cite{nakamura_communication_1987, fabian_baroque_2010}).
    %2) examine whether listeners ``hear'' what our acoustic measurements tell us?
    However, while listeners may be able to identify performers' intentions, they may not have the perceptual acuity to identify certain features with the same precision allowed by acoustic measures. For instance, a study by Howes \cite{howes_relationship_2004} showed there was no correlation between measured and perceived vibrato onset times. 
This suggests that there are some measurable performance parameters that may not map well to human perception.
For example, an objectively measurable difference between a ``deadpan'' and ``expressive'' performance does not necessarily translate to a \textit{perceived}
``expressive'' performance, especially if the changes in measured performance parameters are structurally normative, as discussed in Sect.~\ref{ssec:expr_var}. 
    
Given a weak relation between a measured parameter and listeners' perception of that parameter, an important question arises: is the parameter itself not useful in modeling human perception, or is the metric simply inappropriate?
For example, there are many aspects of music perception that are known to be categorical (e.g., pitch)  in which case a continuous metric would not work well in a model designed to predict human ratings.
Similarly, there is the consideration of the role of the \emph{representation} and transformation of a measured parameter for predicting perceptual ratings.
This question was raised by Timmers, who examined the representation   of tempo and dynamics that best predicted listener judgments of musical similarity \cite{timmers_predicting_2005}. 
This study found that, while most existing models rely on normalized variations of tempo and dynamics, the absolute tempo and the interaction of tempo and loudness were better predictors.

\todo{Clearly, the execution of several performance parameters are important for the perception of both fine-grained and large-grained musical structures, and appear to have
a large influence over listeners' perception and experience of the emotional and expressive aspects of a performance.} %\alex{\pcount is this sentence redundant?} 
% \claire{yes, but I thought that's what conclusions were supposed to do.. I can rewrite it if you need to save space.}
Since the latter appears to carry great significance for both MPA and music perception research, it suggests that future work ought to focus on disentangling the relative weighting of the various features controlled by performers that contribute to an expressive performance. 
Since it is frequently alluded to that a performer's manipulation of musical tension is one of the strongest contributors to an expressive performance, further empirical research must attempt to break down this high-level feature into meaningful collections of well-defined features that would be useful for MPA.

The research surveyed in this section highlights the importance of human perception in MPA research, especially as it pertains to the communication of emotion, musical structure, and creating an aesthetically pleasing performance.
In fact, the successful modeling of perceptually-relevant performance attributes, such as those that mark ``expressiveness,'' could have a large impact not only for MPA
but for many other areas of MIR research, such as computer-generated composition and performance, automatic accompaniment, virtual instrument design and control, or
robotic instruments and HCI (see, e.g., the range of topics discussed in \cite{kirke_guide_2012}).
%Given the already-challenging task of isolating various performance characteristics that contribute to an expressive performance, however, it is clear that both MPA and music perception researchers must ensure that the use of the term ``expressive'' is clearly defined for a given study. 
%Moreover, researchers should be careful not to conflate the various possible usages of the term when interpreting results
A major obstacle impeding research in this area is the inability to successfully isolate (and therefore understand) the various performance characteristics that contribute to a so-called ``expressive'' performance from a listener's perspective. 
Existing literature reviews on the topic of MPA have not been able to shed much light on this problem, in part because researchers frequently disagree (or conflate) the various definitions of ``expressive.''  
%\claire{I'd like to get a citation here? Can't access ``psychology of music'' journal to see Gabrielsson's 2003 review}
Careful experimental design and/or meta-analyses across both MPA and cognition research, as well as cross-collaboration between MIR and music cognition researchers, may therefore prove fruitful areas for future research.  
%\alex{maybe rephrase more in terms of challenges?}
%\claire{better?}